\documentstyle[pra,aps,multicol,epsf]{revtex}
\newcommand{\ket}[1]{| #1 \rangle}
\newcommand{\Ket}[1]{| #1 \rangle _{AB}}

\begin{document}

\draft

\title{Deterministic entanglement concentration}
\author{Fumiaki Morikoshi $^{1,}$ 
\thanks{E-mail address: {\tt fumiaki@will.brl.ntt.co.jp}}
and Masato Koashi $^{2}$}
\address{$^{1}$NTT Basic Research Laboratories, 3-1 Morinosato-Wakamiya,
Atsugi-shi, Kanagawa, 243-0198, Japan\\
$^{2}$CREST Research Team for Interacting Carrier Electronics,
School of Advanced Sciences,\\ The Graduate University for Advanced Studies
(SOKEN), Hayama, Kanagawa, 240-0193, Japan}

\maketitle

\begin{abstract}

Deterministic extraction of Bell pairs from a finite number of partially
entangled pairs is discussed.
We derive the maximum number of Bell pairs that can be obtained with
probability 1 by local operations and classical communication.
It is proved that the optimal deterministic concentration needs only a
two-pair collective manipulation in each step, and that a collective
manipulation of all entangled pairs is not necessary.
Finally, this scheme reveals an entanglement measure for the deterministic
concentration.

\end{abstract}


\begin{multicols}{2}
\narrowtext

\section{Introduction}
\label{Introduction}

Quantum entanglement has come to be viewed as a significant resource for
quantum information processing with the discoveries of quantum teleportation
\cite{Bennett93} and superdense coding \cite{Bennett92}.
A particular entangled state called a Bell pair,
\begin{equation}
 \Ket{\Phi^{+}} = \frac{1}{\sqrt{2}} (\Ket{00} + \Ket{11}),
\end{equation}
is an essential ingredient for these communications, where qubits $A$ and $B$
are possessed by Alice (sender) and Bob (receiver), respectively.
It is thus important for Alice and Bob to prepare Bell pairs between
them in advance, which is related to the main result in this paper.

In order to seek new applications and efficient manipulation of entangled
states, we need a deep understanding of the nature of quantum entanglement.
Most recent investigations of entangled states have been undertaken
within the framework of local operations and classical communication (LOCC)
(for a review, see Ref. \cite{Plenio}).
Since Alice and Bob are supposed to be separated by a large distance,
they are allowed to access only their own systems locally, and to communicate
in a classical manner.
In other words, neither global operations on the whole system nor
transmission of quantum systems is allowed.
Thus the following question is crucial for a good grasp of entanglement:
What can we do on entangled states by using LOCC?

In this paper, we deal with only bipartite pure entangled states, for which
general theorems useful in answering the above question have already been
proved \cite{Nielsen,Vidal99,Jonathan99-1}.
Nielsen proved the necessary and sufficient conditions for deterministic
transformations between bipartite pure entangled states \cite{Nielsen}.
In Ref. \cite{Vidal99}, Vidal derived the formula for the maximum probability
with which a pure entangled state is transformed into another one.
Jonathan and Plenio extended Nielsen's theorem to the case where the final
state is an ensemble of pure states \cite{Jonathan99-1}.

Transformations whose final states include Bell pairs are of special
interest among possible entanglement transformations, because the resultant
Bell pairs can be used for quantum communication such as teleportation.
Extraction of maximally entangled states from partially entangled states
by LOCC is generally called entanglement concentration.
Various studies on its limitations and efficiency have been carried out
\cite{Vidal99,Jonathan99-1,Bennett96,Lo,Bose}.
First, Bennett {\it et al.} constructed an entanglement concentration
procedure, which extracts Bell pairs from many identical copies of a
partially entangled pair \cite{Bennett96}.
Its efficiency in the asymptotic limit is characterized by the von Neumann
entropy of Alice's (or Bob's) reduced density matrix.
This fact, together with the existence of the reverse procedure,
called entanglement dilution, established von Neumann entropy as
a unique measure of pure-state entanglement in the asymptotic limit
\cite{Popescu}.
In Refs. \cite{Vidal99,Lo}, Lo and Popescu gave a formula for the maximum
probability of obtaining a maximally entangled state from a partially
entangled state (and Vidal later extended it for more general final states).
Furthermore, Jonathan and Plenio optimized the average amount of entanglement
obtained in the entanglement concentration that transforms a partially
entangled state into an ensemble of maximally entangled states in various
dimensions \cite{Jonathan99-1}.
All these entanglement concentration processes are probabilistic for a finite
number of entangled pairs.

The entanglement concentration scheme presented in this paper is designed
for a somewhat different purpose.
It deterministically extracts Bell pairs from two-qubit entangled pairs.
In other words, we focus on a concentration that converts a collection of
two-qubit partially entangled pairs having different amounts of entanglement
into a bunch of Bell pairs with probability 1.
While the above probabilistic concentration processes run the risk of losing
all the entanglement contained in the initial states with certain
probability, our concentration scheme answers the question of how many Bell
pairs can be obtained from partially entangled pairs without gambling.
On the other hand, it is probable in practical applications that Alice and
Bob share many partially entangled pairs with different entanglement, and
wish to prepare Bell pairs from them.
Thus we consider transformations whose initial states are tensor products
of partially entangled pairs.

This paper derives the maximum number of Bell pairs obtained in the
deterministic entanglement concentration, and proves that the optimal
procedure needs only a two-pair entanglement manipulation in each step.
This contrasts sharply with the fact that the probabilistic concentrations
known so far requires a collective manipulation of all entangled pairs for
optimality.
Furthermore, this scheme reveals a reasonable entanglement measure for the
deterministic concentration.

This paper is organized as follows.
We begin in Sec. \ref{Transfer} by introducing the main tool used
in this paper, that is, Nielsen's theorem.
Then we review the transfer of entanglement between two partially entangled
pairs in a slightly different way from Ref. \cite{Morikoshi}.
The argument is essentially the deterministic entanglement concentration of
two pairs.
In Sec. \ref{Deterministic}, we deal with the deterministic concentration
of $n$ pairs, which is the main result of this work.
The maximum number of Bell pairs is derived and proved to be attained
by a series of collective manipulations of two pairs.
Section \ref{Measure} introduces an entanglement measure for the
deterministic concentration.
Section \ref{Schematic} describes a schematic representation of the
concentration scheme.
Then we extend the entanglement measure to general pure entangled states in
Sec. \ref{General}.
Finally, Sec. \ref{Conclusion} concludes the paper.

\section{Transfer of entanglement between two pairs}
\label{Transfer}

In this section, we treat the transfer of entanglement between two pairs,
which is the basis of the entanglement concentration of $n$ pairs.
This subject was already examined from a different point of view,
the recovery of entanglement, in Ref. \cite{Morikoshi}.
When Alice and Bob transform an entangled state into another one by LOCC,
the quantity of entanglement generally decreases.
However, the entanglement lost in the manipulation can be partially
recovered by an auxiliary entangled pair.
Attaching an auxiliary entangled pair, Alice and Bob can partially transfer
the entanglement lost in the original pair to the auxiliary one
by collective manipulation of both pairs.
In the following, the result from the recovery scheme is briefly reviewed
in a useful form for entanglement concentration.

First we introduce Nielsen's theorem and its mathematical basis.
The theorem is described by the mathematical theory of majorization,
which gives an ordering of real vectors according to their degree of disorder
\cite{Bhatia}.
Let $x=(x_1, \ldots, x_n)$ and $y=(y_1, \ldots, y_n)$ be
real $n$-dimensional vectors.
Rearranging the elements of $x$ in decreasing order,
we obtain $x^{\downarrow}=(x^{\downarrow}_1, \ldots, x^{\downarrow}_n)$,
where $x^{\downarrow}_1 \geq \cdots \geq x^{\downarrow}_n$.
We say that $x$ is majorized by $y$, written as $x \prec y$, if
\begin{equation}
 \sum_{j=1}^{k} x^{\downarrow}_{j} \leq \sum_{j=1}^{k} y^{\downarrow}_{j},
 \qquad 1 \leq k \leq n-1,
\end{equation}
for $k=1, \ldots, n-1$, with equality instead of inequality for $k=n$.

Any bipartite pure entangled states can be written in a standard form called
Schmidt decomposition:
$\Ket{\psi} = \sum _{i} \sqrt{a_{i}} |i\rangle_{A} |i\rangle_{B}$,
where $\{|i\rangle_{A}\}$ and $\{|i\rangle_{B} \}$ are the orthonormal
bases of respective systems.
The eigenvalues of the reduced density matrix
$\rho_{\psi} \equiv tr_{B}(|\psi \rangle _{AB \, AB} \langle \psi|)$ are
$a_{1}, \ldots, a_{n}$, which are positive real numbers, and sum to 1.
We define the vector of these eigenvalues as
$\lambda_{\psi} \equiv (a_{1}, \ldots, a_{n})$.
With these definitions and the theory of majorization, Nielsen proved that
a bipartite pure entangled state $\Ket{\psi}$ is transformed into another one
$\Ket{\phi}$ with probability 1 by LOCC, if and only if
$\lambda_{\psi}$ is majorized by $\lambda_{\phi}$, i.e.,
\begin{equation}
 \Ket{\psi} \to \Ket{\phi} \quad \text{iff}
 \quad \lambda_{\psi} \prec \lambda_{\phi}.
\end{equation}

One way of quantifying entanglement is to use the von Neumann entropy of
a reduced density matrix:
\begin{equation}
 E(\psi) \equiv - \mathrm{tr} (\rho_{\psi} \log_{2} \rho_{\psi}).
\label{eq:entropy}
\end{equation}
This measure is related to the efficiency with which an infinite number of
identical copies of a partially entangled pair are concentrated
\cite{Bennett96}.
It is known that entanglement generally decreases during deterministic
transformations by LOCC, i.e., if
\begin{equation}
 \Ket{\psi} \to \Ket{\phi} \quad \text{then} \quad E(\psi) \geq E(\phi).
\end{equation}
Therefore, when entanglement is transferred from an entangled pair to another
one, the total entanglement of two pairs generally decreases.

Next we proceed to the transfer of entanglement between two pairs.
The process can be considered as the entanglement concentration of two pairs
if the purpose is extraction of a Bell pair.
Suppose Alice and Bob share two partially entangled pairs
\begin{equation}
\begin{array}{rcl}
 \ket{\psi} &=& \sqrt{a} \ \ket{00} + \sqrt{1-a} \ \ket{11}, \\
 \ket{\phi} &=& \sqrt{b} \ \ket{00} + \sqrt{1-b} \ \ket{11}. \\
\end{array}
\end{equation}
Without loss of generality, we can set
\begin{equation}
 \frac{1}{2} < a \leq b < 1.
 \label{eq:a}
\end{equation}
According to Nielsen's theorem, neither of these two pairs can be
concentrated to a Bell pair with probability 1.
However, when $ab \leq \frac{1}{2}$, it is possible that Alice and Bob
extract a Bell pair from the combined entangled state
$\ket{\psi} \otimes \ket{\phi}$ with probability 1, as shown below.

Case (a): $\frac{1}{4} < ab \leq \frac{1}{2}$.
In this case, Alice and Bob can extract a Bell pair from
$\ket{\psi} \otimes \ket{\phi}$ and keep the residual entanglement in another
entangled pair $\ket{\omega}$.
The state that maximizes the residual entanglement is
\begin{equation}
 \ket{\omega} = \sqrt{2ab} \ \ket{00} + \sqrt{1-2ab} \ \ket{11}. \\
\label{eq:residual1}
\end{equation}
This concentration is proved to be possible by Nielsen's theorem as follows.
The vectors of eigenvalues for the initial and the final states are
\begin{equation}
 \begin{array}{l}
 \lambda_{\psi \otimes \phi} =
 \left( \begin{array}{c} a \\ 1-a \end{array} \right) \otimes
 \left( \begin{array}{c} b \\ 1-b \end{array} \right)
 = \left( \begin{array}{c} ab \\ a(1-b) \\ (1-a)b \\ (1-a)(1-b)
 \end{array} \right), \\
  \lambda_{\Phi^{+} \otimes \omega} =
 \left( \begin{array}{c} \frac{1}{2} \\ \frac{1}{2} \end{array} \right)
 \otimes
 \left( \begin{array}{c} 2ab \\ 1-2ab \end{array} \right)
 = \left( \begin{array}{c} ab \\ \frac{1}{2}-ab \\ ab \\ \frac{1}{2}-ab
 \end{array} \right),
 \end{array}
\label{eq:transfer1}
\end{equation}
respectively.
In order to rearrange the elements of these vectors in decreasing order,
all we have to do is just reverse the order of the second and third
elements of each vector because of Eq. (\ref{eq:a}).
Since Eq. (\ref{eq:a}) guarantees the inequalities
\begin{equation}
 b \leq 2ab, \quad (1-a)(1-b) \geq \frac{1}{2} + ab,
\end{equation}
the majorization relation
$\lambda_{\psi \otimes \phi} \prec \lambda_{\Phi^{+} \otimes \omega}$ holds.
Thus Alice and Bob can perform the concentration
$\ket{\psi} \otimes \ket{\phi} \to \ket{\Phi^{+}} \otimes \ket{\omega}$.
If the larger amplitude of $\ket{\omega}$, $\sqrt{2ab}$, becomes smaller,
the first inequality for $\lambda_{\psi \otimes \phi} \prec
\lambda_{\Phi^{+} \otimes \omega}$ does not hold.
Therefore, Eq. (\ref{eq:residual1}) is optimal in the sense that it contains
as much residual entanglement as possible.

Case (b): $\frac{1}{2} < ab < 1 $.
It is impossible for Alice and Bob to extract a Bell pair in a deterministic
way.
The best they can do is gather entanglement of two pairs into one pair
to make an entangled pair as close to a Bell pair as possible.
That is, they perform
$\ket{\psi} \otimes \ket{\phi} \to \ket{\omega} \otimes \ket{00}$, where
\begin{equation}
 \ket{\omega} = \sqrt{ab} \ \ket{00} + \sqrt{1-ab} \ \ket{11}. \\
\end{equation}
We can prove this transformation to be possible by Nielsen's theorem again.
The vector of eigenvalues for the final state becomes
\begin{equation}
 \begin{array}{rcccl}
 \lambda_{\omega \otimes \ket{00}} &=&
 \left( \begin{array}{c} ab \\ 1-ab \end{array} \right) \otimes
 \left( \begin{array}{c} 1 \\ 0 \end{array} \right) &=&
 \left( \begin{array}{c} ab \\ 0 \\ 1-ab \\ 0
 \end{array} \right). \\
 \end{array}
\end{equation}
Clearly, this vector majorizes the vector of the initial state
$\lambda_{\psi \otimes \phi}$ [Eq. (\ref{eq:transfer1})].
Thus Alice and Bob succeed in collecting the entanglement of
$\ket{\psi}$ and $\ket{\phi}$ into $\ket{\omega}$,
whose entanglement is maximized for the same reason as the previous case.

\begin{figure}
\begin{center}
\epsfxsize=5cm
\epsfbox{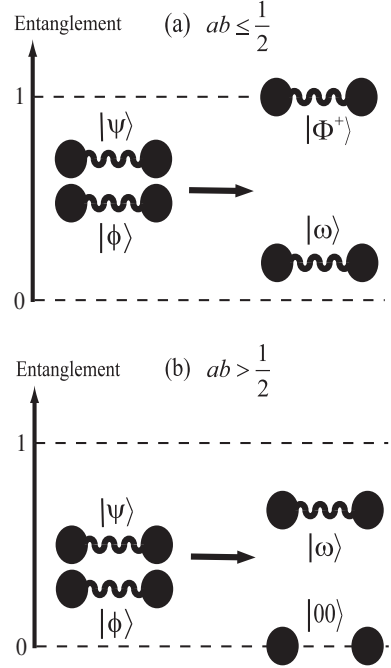}
\end{center}
\caption{Optimal deterministic concentration of two pairs
$\ket{\psi}=\sqrt{a}\ket{00}+\sqrt{1-a}\ket{11}$ and
$\ket{\phi}=\sqrt{b}\ket{00}+\sqrt{1-b}\ket{11}$.
Each pair connected by a wavy line represents an entangled pair.
The axis indicates the quantity of entanglement.
(a) When $ab \leq \frac{1}{2}$, Alice and Bob can extract a Bell pair
$\ket{\Phi^{+}}$, and keep a residual entangled pair $\ket{\omega}$.
(b) When $ab>\frac{1}{2}$, they gather all the entanglement to
$\ket{\omega}$, which is not maximally entangled.}
\label{Fig1}
\end{figure}

These two cases of deterministic concentration of two pairs are
illustrated in Fig. \ref{Fig1}.
In both cases, the most important point is that
the largest element of the four-dimensional vector, $ab$, never changes
during the concentration process.
In other words, the product of the larger amplitudes of each entangled pair
is conserved before and after the transformation, which is the crux of
the optimality of the deterministic concentration presented in Sec.
\ref{Deterministic}.

\section{Deterministic entanglement concentration of \lowercase{$n$} pairs}
\label{Deterministic}

Now we move on to the main topic of this paper: the concentration of a finite
number of partially entangled pairs.
Suppose Alice and Bob share $n$ partially entangled pairs
$\ket{\psi_{1}} \otimes \cdots \otimes \ket{\psi_{n}}$,
where
\begin{equation}
 \ket{\psi_{i}} = \sqrt{a_{i}} \ \ket{00} + \sqrt{1-a_{i}} \ \ket{11}
 \quad (i=1, \ldots, n),
\label{eq:partially}
\end{equation}
with
\begin{equation}
 \frac{1}{2} < a_{i} < 1,
\end{equation}
and wish to extract as many Bell pairs as possible from these entangled
pairs with probability 1.
The final state is $ \otimes ^{k} \ket{\Phi^{+}} \otimes ^{n-k} \ket{00}$
if they obtain $k$ Bell pairs and $(n-k)$ disentangled pairs after the
concentration.
Thus the vectors corresponding to the initial and the final states are
$ \left( \begin{array}{c} a_{1} \\ 1-a_{1} \end{array} \right) \otimes
 \cdots \otimes \left( \begin{array}{c} a_{n} \\ 1-a_{n} \end{array} \right)$
and
$ \otimes ^{n-k} \left( \begin{array}{c} 1 \\ 0 \end{array} \right)
 \otimes ^{k} \left( \begin{array}{c} \frac{1}{2} \\ \frac{1}{2} \end{array}
 \right) $,
respectively.

This concentration is accomplished if the following majorization relation
holds:
\begin{equation}
 \left( \begin{array}{c} a_{1} \cdots a_{n} \\ \vdots \\ \vdots \\ \vdots \\
 (1-a_{1}) \cdots (1-a_{n}) \end{array} \right) \prec
 \left( \begin{array}{c} \frac{1}{2^{k}} \\ \vdots \\ \frac{1}{2^{k}} \\
 0 \\ \vdots \\ 0 \end{array} \right). 
\label{eq:concentration}
\end{equation}

In the initial state, the first element of the vector is the largest one.
All the nonzero elements of the vector for the final state are equal to
$2^{-k}$.
Therefore, if the first inequality of majorization relation
(\ref{eq:concentration}) holds, i.e.,
\begin{equation}
 a_{1} \cdots a_{n} \leq \frac{1}{2^{k}},
\label{eq:first}
\end{equation}
this concentration is successfully completed
because the other inequalities for Eq. (\ref{eq:concentration})
are automatically satisfied.

Inequality (\ref{eq:first}) gives the maximum number of Bell pairs that can
be extracted from the partially entangled pairs given by
Eq. (\ref{eq:partially}),
\begin{equation}
 k_{max} = \lfloor -\log_{2} (a_{1} \cdots a_{n}) \rfloor,
\label{eq:maximum}
\end{equation}
where $\lfloor x \rfloor$ represents the largest integer equal to or
less than $x$.

Equation (\ref{eq:maximum}) tells us that the maximum number of Bell pairs
is determined only by the product of the parameters of each pair to be
concentrated, i.e., $a_{1} \cdots a_{n}$.
This fact guarantees that we lose no Bell pairs as long as the product is
conserved.
In Sec. \ref{Transfer}, we found that the product never changes
during the optimal deterministic concentration of two pairs.
This means that we can attain the maximum number of Bell pairs through
repeated use of two-pair concentrations.

The actual concentration procedure is as follows.
Alice and Bob choose two arbitrary entangled pairs, then perform
a concentration on them.
If they obtain a Bell pair [the case in Fig. \ref{Fig1}(a)],
it is put aside, and the residual entangled pair will be reused in the next
stage of concentration.
If the selected two pairs are not enough for a Bell pair
[the case in Fig. \ref{Fig1}(b)],
they transfer all the entanglement contained in a pair into another one
to make a partially entangled pair that will be reused in the next pairwise
concentration.
They repeat these operations until they achieve the maximum number of Bell
pairs.

By repeating the two-pair concentrations as stated above, Alice and Bob
finally obtain $k_{max}$ Bell pairs, one partially entangled pair, and
$(n-k_{max}-1)$ disentangled pairs.
The product of the larger elements of vectors for each pair never changes
throughout these concentration processes, i.e., it is always
$a_1 \cdots a_n$.
Therefore, the residual partially entangled state becomes
\begin{equation}
 \ket{\omega} = \sqrt{2^{k_{max}}a_{1} \cdots a_{n}} \ket{00} +
 \sqrt{1-2^{k_{max}}a_{1} \cdots a_{n}} \ket{11},
\label{eq:residual2}
\end{equation}
where $\frac{1}{2} < 2^{k_{max}}a_{1} \cdots a_{n} \leq 1$,
because there are $k_{max}$ Bell pairs and $(n - k_{max} -1)$ disentangled
pairs besides that partially entangled pair.
The amount of entanglement of the residual pair cannot exceed that of
Eq. (\ref{eq:residual2}) for the same reason as stated in the proof of the
two-pair concentration.

The product $a_{1} \cdots a_{n}$ is not usually equal to $2^{-k_{max}}$.
Thus $k_{max}$ Bell pairs do not represent all the available entanglement.
By preserving the residual entanglement in Eq. (\ref{eq:residual2}) and
conserving the quantity $a_{1} \cdots a_{n}$,
Alice and Bob do not waste any potential entanglement that can  be extracted
in a future concentration, which attains the true optimality
in this deterministic concentration.
That is, if they obtain an extra partially entangled pair later, they will be
able to perform a pairwise concentration on the residual pair $\ket{\omega}$
and the extra pair.

Next we consider the complexity of the deterministic concentration.
In each pairwise concentration, at least one of the pairs becomes a Bell pair
or a disentangled pair, which does not proceed to the next concentration.
Thus, if we perform a pairwise concentration at most $n-1$ times in a
process similar to elimination in a tournament, there finally remain
$k_{max}$ Bell pairs, the residual pair $\ket{\omega}$, and $(n-k_{max}-1)$
disentangled pairs.

According to the prescription given in Ref. \cite{Nielsen}, a transformation
of an entangled state with Schmidt number $m$ can be performed by $m-1$
operations, where each operation consists of transformations in the
two-dimensional subspace of the entire entangled state.
Thus, when we collectively manipulate $n$ entangled pairs consisting
of two qubits, i.e., an entangled state with Schmidt number $2^n$,
we generally perform $O(2^n)$ operations.
On the other hand, the number of steps in the deterministic concentration of
$n$ pairs is reduced to $O(n)$ by pairwise manipulation, because each
pairwise concentration needs four operations.
This argument suggests the importance of the complexity of entanglement
manipulation.
It is worth considering not only what we can do on entangled states,
but also how efficiently we can manipulate them.

The pairwise nature of this concentration scheme is useful in a practical
sense too.
Even if we fail in the manipulation of two pairs, or if some error occurs
in the concentration, only the two pairs are affected.
The other pairs remain intact in spite of such unexpected effects.
This contrasts sharply with other concentration schemes that require
collective manipulation of all pairs.

\begin{figure}
\begin{center}
\epsfxsize=5.5cm
\epsfbox{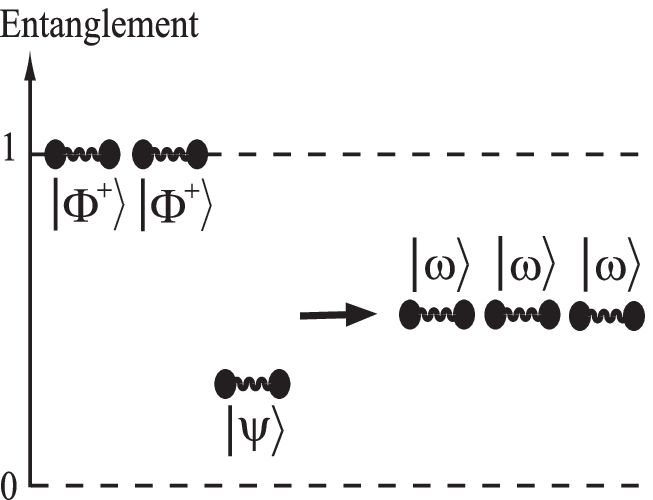}
\end{center}
\caption{The transformation
$\ket{\Phi^{+}} \otimes \ket{\Phi^{+}} \otimes \ket{\psi} \to$
$\ket{\omega} \otimes \ket{\omega} \otimes \ket{\omega}$
requires a collective manipulation of the three pairs,
where $\ket{\Phi^{+}}$ represents a Bell pair,
$\ket{\psi}=\sqrt{0.95}\ket{00}+\sqrt{0.05}\ket{11}$, and
$\ket{\omega}=\sqrt{0.9}\ket{00}+\sqrt{0.1}\ket{11}$.
Two-pair manipulation cannot pump up $\ket{\psi}$.}
\label{Fig2}
\end{figure}

In general, collective manipulation of more than two pairs is necessary
even for transformations whose initial and final states consist of 
tensor products of entangled pairs.
For example, the transformation illustrated in Fig. \ref{Fig2} requires
a collective manipulation of three pairs.
The vectors for the initial and the final state are
$(0.5, 0.5) \otimes (0.5, 0.5) \otimes (0.95, 0.05)$ and
$(0.9, 0.1) \otimes (0.9, 0.1) \otimes (0.9, 0.1)$, respectively.
These tensor products are calculated in decreasing order as follows:
\linebreak[4]
$\lambda_{initial} = (0.2375, 0.2375, 0.2375, 0.2375, 0.0125, 0.0125,
\linebreak[4] 0.0125, 0.0125)$ and
$\lambda_{final} = (0.729, 0.081, 0.081, 0.081, \linebreak[4] 0.009,
0.009, 0.009, 0.001)$.
It is easily seen that $\lambda_{initial} \prec \lambda_{final}$,
and thus this transformation is really possible according to Nielsen's
theorem.
However, it requires collective manipulation of three pairs because
it was proved that two-pair manipulation cannot increase the entanglement
of the less entangled pair in the initial state \cite{Morikoshi}.
That is, any successive operations of two pairs cannot pump up the lowest
state $(0.95, 0.05)$.
As shown in this example, not all transformations of entangled pairs consist
of two-pair manipulations.
Therefore, it is remarkable that two-pair manipulation is enough for the
deterministic concentration.

\section{Entanglement measure for the deterministic concentration}
\label{Measure}

The argument in Sec. \ref{Deterministic} leads to an entanglement measure for
the deterministic concentration of a finite number of pairs.
First we present conditions for entanglement measures in deterministic
transformations of pure states, which is a modified version of the conditions
for entanglement measures proposed in Refs. \cite{Vedral97,Vedral98}.
When we restrict ourselves to manipulating entangled pure states in a
deterministic manner, a measure of entanglement $M(\psi)$ should satisfy the
following conditions:
\begin{list}{}{}
 \item (i) $M(\psi)=0$ iff $\ket{\psi}$ is separable.
 \item (ii) $M(\psi)$ remains unchanged under local unitary transformations.
 \item (iii) $M(\psi)$ cannot be increased by deterministic LOCC.
\end{list}
Due to the restriction of deterministic transformations, the third condition
is weaker than the counterpart in Refs. \cite{Vedral97,Vedral98}, which
requires the nonincreasing property of the {\it expected} value of the
measure when the final states are not unique.

Now we derive an entanglement measure for deterministic concentration.
The formula for the maximum number of Bell pairs [Eq. (\ref{eq:maximum})],
can also be written in the following form:
\begin{equation}
 k_{max} = \left\lfloor \sum _{i=1} ^{n} - \log_{2} a_{i} \right\rfloor
\end{equation}
This expression implies that the number of Bell pairs that a partially
entangled state
$\ket{\psi_{i}} = \sqrt{a_{i}} \ \ket{00} + \sqrt{1-a_{i}} \ \ket{11}$
contains is
\begin{equation}
 D(\psi_{i}) \equiv - \log_{2} a_{i} \qquad (\frac{1}{2} \leq a_{i} \leq 1).
\label{eq:measure}
\end{equation}
It is easily seen that this quantity satisfies the above conditions for
entanglement measures in deterministic cases.
If we allow probabilistic transformations, the expected value of our measure
can be increased in some cases.
Note that the measure $D(\psi_{i})$ does not diverge since it is defined
with the square of the larger Schmidt coefficient of $\ket{\psi_{i}}$,
which ranges from $\frac{1}{2}$ to $1$.
In addition, a Bell pair corresponds to the unit of this measure, i.e.,
$D(\ket{\Phi^{+}}) = 1$.

Another entanglement measure $E(\psi)$ defined by Eq. ({\ref{eq:entropy}})
represents the amount of entanglement per one partially
entangled pair $\ket{\psi}$ in the asymptotic limit.
This means that if we have an infinite number of identical copies of
$\ket{\psi}$, we can extract $E(\psi)$ Bell pairs per copy.
On the other hand, $D(\psi_{i})$ quantifies the number of Bell pairs
contributed by the state $\ket{\psi_{i}}$ when we deterministically
concentrate a finite number of entangled pairs that are not generally
identical.
For example, if Alice and Bob share two pairs $\ket{\psi_{1}}$ and
$\ket{\psi_{2}}$, where $D(\psi_{1})=0.7$ and $D(\psi_{2})=0.6$,
then they can obtain a Bell pair and a residual state $\ket{\omega}$
with $D(\omega)=0.3$, because $0.6+0.7=1+0.3$.

Since $D(\psi)$ is smaller than $E(\psi)$, our concentration scheme is less
efficient than the original Schmidt projection method \cite{Bennett96},
which attains $E(\psi)$ in the asymptotic limit.
The inefficiency is due to the strong restriction of deterministic
transformations.
Thus the quantity $D(\psi)$ does not converge on $E(\psi)$ even in the
asymptotic limit.

The most fascinating property of this measure $D(\psi)$ is that
the maximum number of Bell pairs is determined by the addition of the
contribution of each pair.
This property shows that the deterministic concentration procedure does not
depend on the way of pairing entangled pairs, because we can attain the
optimality as long as the sum $\sum_{i} D(\psi_{i})$ does not decrease. 
This fact assures the optimality of deterministic concentration  by {\it any}
pairwise manipulation.

\section{Schematic representation of the deterministic concentration}
\label{Schematic}

\begin{figure}
\begin{center}
\epsfxsize=5cm
\epsfbox{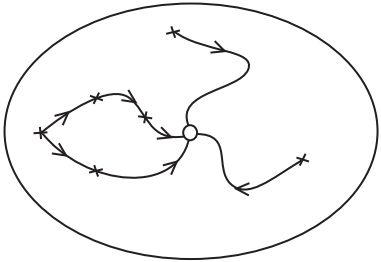}
\end{center}
\caption{The elements of this set are tensor products of two-qubit
entangled pairs, and have the same value of $\sum_{i} D(\psi_{i})$.
The center represents a bunch of the maximum number of Bell pairs and
a residual entangled pair, i.e., {\it the drain}.
Every point (x) in this set is led to the drain by any transformation that
reduces the total amount of von Neumann entropy $\sum_{i} E(\psi_{i})$.}
\label{Fig3}
\end{figure}

In this section, we schematically represent our understanding of the
deterministic concentration.
Figure \ref{Fig3} represents a set of entangled states such that each
element is a tensor product of two-qubit entangled pairs, and has the same
value of $\sum_{i} D(\psi_{i})$.
The center of the set is the point that represents a bunch of $k_{max}$ Bell
pairs and a residual entangled pair.
As shown in Sec. \ref{Measure}, all points in the set can be transformed
into the center.
In other words, wherever the initial state is, every path surprisingly leads
to the center as long as the transformation conserves the sum of
$D(\psi_{i})$.
The center representing $k_{max}$ Bell pairs and a residual pair acts like
a {\it drain}.

The advantage of the deterministic concentration is explained schematically
as follows.
Since pairwise manipulation is enough for optimality,
any point representing a collection of partially entangled pairs are
transformed into the center along the path that changes only two pairs
at a time.
Furthermore, our concentration procedure does not depend on the way of
pairing entangled pairs.
This corresponds to the fact that there are many possible paths that
connect an initial state and the drain.

Figure \ref{Fig3} is also explained in terms of the entropy of entanglement
$E(\psi)$ and the entanglement measure $D(\psi)$.
The total entropy generally decreases in deterministic transformations.
Thus Fig. \ref{Fig3} means that as long as $\sum_{i} D(\psi_{i})$ is
conserved, every point in the set is always led to the center by any
transformation that decreases the total entropy.
The drain that represents $k_{max}$ Bell pairs and a residual pair is the
most stable state under the condition that the sum of $D(\psi_{i})$ remains
unchanged.
In other words, whatever entangled pairs Alice and Bob share, those pairs are
swept up to a bunch of Bell pairs and a bunch of disentangled pairs.
All we need for the optimality of deterministic concentration is the
conservation of  $\sum_{i}D(\psi_{i})$.

\section{Extension to general pure entangled states}
\label{General}

The entanglement measure for the deterministic concentration $D(\psi)$,
defined in Eq. (\ref{eq:measure}), can be extended to general pure entangled
states.
Suppose Alice and Bob share a partially entangled state whose Schmidt number
is $m$,
\begin{equation}
 \ket{\psi} = \sum_{j=1}^{m} \sqrt{p_{j}} \ \ket{j,j} \qquad (m \geq 2),
\end{equation}
where
\begin{equation}
 \sum_{j=1}^{m} p_{j}=1.
\end{equation}
By the same argument as in Sec. \ref{Deterministic},
we can determine the maximum number of Bell pairs extractable from this
partially entangled state:
\begin{equation}
 k_{max} = \left\lfloor -\log_{2} (\max\limits_{i} p_{i}) \right\rfloor.
\end{equation}
Thus the contribution of the partially entangled state $\ket{\psi}$ to the
maximum number of Bell pairs is also expressed as
\begin{equation}
 D(\psi) \equiv - \log_{2} (\max\limits_{i} p_{i})
 \qquad (0 < \max\limits_{i} p_{i} \leq 1).
\label{eq:measure2}
\end{equation}
Of course, Eq. (\ref{eq:measure2}) satisfies the conditions for the
entanglement measure in deterministic cases, as in Sec. \ref{Measure}.

From Eq. (\ref{eq:measure2}), it is easily seen that $D(\psi)$ is additive
for a system composed of independent subsystems, namely,
\begin{equation}
 D(\psi \otimes \phi) = D(\psi) + D(\phi),
\end{equation}
because the largest element of the vector for $\ket{\psi} \otimes \ket{\phi}$
is the product of those for $\ket{\psi}$ and $\ket{\phi}$.
We also point out that both $D(\psi)$ and another additive
quantity $E(\psi)$ are obtained by some limits of a quantity known as Renyi
entropy of order $t$,
$H_{t} (\{p_{i} \}) = [1/(1-t)] \log_{2} \sum_{i=1}^{n} p_{i}^t$,
which is also additive.
That is, $H_{1} (\{ p_{i} \}) = E(\{ p_{i} \})$ and $H_{\infty} (\{ p_{i} \})
= D(\{ p_{i} \})$,
where $p_{i}$ represents the square of each Schmidt coefficient.

Though we succeeded in extending the entanglement measure to general cases,
the actual concentration procedures are different from
that of two-qubit cases in some points.
First, the residual entanglement represented by $\{D(\psi) - k_{max} \}$
cannot be always kept in a partially entangled pair.
A simple example is the transformation from a maximally entangled pair with
Schmidt number three to a Bell pair.
Since $\lfloor \log_{2} 3 \rfloor =1$, we can extract a Bell pair from 
$(1/\sqrt{3}) (\ket{00}+\ket{11}+\ket{22})$.
If the residual entanglement represented by
$(\log_{2} 3  -1)$ is kept in another partially entangled
pair, the Schmidt number of the final state becomes 4, which is impossible
because we cannot increase the Schmidt number of initial states.
This also means the optimal concentration by tournamentlike operations is
impossible in general cases, because it is necessary for the optimal
tournamentlike concentration to retain the residual entanglement in the
form of a partially entangled pair in each step.

\begin{figure}
\begin{center}
\epsfxsize=5cm
\epsfbox{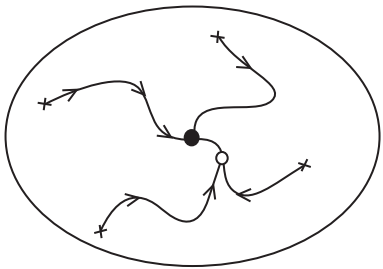}
\end{center}
\caption{The elements of this set are general pure entangled states that have
the same value of $D(\psi)$.
All the elements are led to the center, {\it the drain}, by any
transformation that reduces the von Neumann entropy $E(\psi)$.
The symbol O represents a bunch of the maximum number of Bell pairs and a
residual entangled pair.
In general, this state is different from the drain, because not all the
elements are led to the state.}
\label{Fig4}
\end{figure}

Second, the schematic representation of the generalized cases is slightly
different from that of the two-qubit case.
When the initial state is a collection of two-qubit entangled pairs,
every state can be transformed into the state with a bunch of Bell pairs and
a residual entangled pair, as shown in Sec. \ref{Schematic}.
However, in general cases, not all initial states reach that state.
All states with the same value of $D(\psi)$ are transformed into the
following state, which is  similar to the $d$-dimensional maximally
entangled state, where $d = \lfloor 1/(\max\limits_{i} p_{i}) \rfloor$.
(for simplicity, we use vector notation):
\begin{equation}
 (\overbrace{\max\limits_{i} p_{i},\  \cdots,\  \max\limits_{i} p_{i}}^{d},\ 
  1- d \ (\max\limits_{i} p_{i}),\  0, \ \cdots,\  0).
\end{equation}
This state is the drain in the generalized cases.
As shown in Fig. \ref{Fig4}, the point representing a bunch of Bell pairs
and a residual entangled pair is different from the drain in the general case.
[When $D(\psi)$ is equal to an integer, two points coincide because there is
no residual entanglement.]
Of course, we can always reach the state of $k_{max}$ Bell pairs from the
drain, though the state is not in the set shown in Fig. \ref{Fig4}.
The deterministic concentration of two-qubit entangled pairs is a
special case, which always enables us to retain the residual entanglement in
a compact form of a partially entangled pair.

With this generalization of $D(\psi)$, we can prove that entanglement
catalysis \cite{Jonathan99-2} cannot enhance the efficiency of the
deterministic entanglement concentration.
Since an attached catalyst state $\ket{\phi}$ must remain intact after the
transformation, the term $D(\phi)$ does not contribute to the number of
extractable Bell pairs due to the additivity of $D(\psi)$.
Thus the concentration scheme presented in this paper is the best we can do
in deterministic cases.

\section{Conclusion}
\label{Conclusion}

We have presented a deterministic entanglement concentration scheme for
a finite number of partially entangled pairs consisting of two qubits.
We determined the maximum number of Bell pairs that can be extracted by LOCC,
which was proved to be attained by a series of two-pair manipulation
in each step.
There is no need of collective manipulation of all entangled pairs for
deterministic concentration.
The complexity of entanglement manipulation also deserves further
investigation for a deeper understanding of quantum resources.

Furthermore, this concentration scheme revealed an entanglement measure
for the deterministic concentration of a finite number of pairs.
The measure represents the amount of entanglement that we can use without
fail.
In the concentration of two-qubit entangled pairs, we proved that all the
entanglement quantified by the measure can be used without loss.
However, in the general cases, what kind of initial states can be
concentrated into Bell pairs without discarding any residual entanglement
remains open.

\section*{Acknowledgments}

We are grateful to N. Imoto for valuable discussions, and for a careful
reading of the manuscript.
This work was supported by a Grant-in-Aid for Encouragement of Young
Scientists (Grant No.~12740243) and a Grant-in-Aid for Scientific Research
(B) (Grant No.~12440111) by the Japan Society of the Promotion of Science.

\end{multicols}

\end{document}